\documentclass[%
 aip,
 amsmath,amssymb,
preprint,%
]{revtex4-1}

\usepackage{graphicx}
\usepackage{dcolumn}
\usepackage{bm}
\usepackage[utf8]{inputenc}
\usepackage[T1]{fontenc}
\usepackage{mathptmx}
\usepackage{xcolor}
\usepackage{url}
\usepackage{hyperref}
\usepackage[normalem]{ulem}
\usepackage{subfigure}

\hypersetup{
 colorlinks = true,
 allbordercolors = {white},
 allcolors = {blue},
 }
\usepackage[symbol]{footmisc}

\begin{document}

\title{Perturbation Field in The Presence of Uniform Mean Flow: Doppler Effect for Flows and Acoustics}

\author{Tapan K. Sengupta\footnote{Corresponding author.}}
\email{tksengupta@iitism.ac.in}
\affiliation{Department of Mechanical Engineering, IIT (ISM) Dhanbad, Jharkhand-826 004, India}

\author{Aditi Sengupta}
\affiliation{Department of Mechanical Engineering, IIT (ISM) Dhanbad, Jharkhand-826 004, India}

\author{Bhavna Joshi}
\affiliation{Department of Mechanical Engineering, IIT (ISM) Dhanbad, Jharkhand-826 004, India}

\author{Prasannabalaji Sundaram}
\affiliation{CERFACS, Toulouse, France}


 


\date{\today}

\begin{abstract}
Having developed the perturbation equation for a  dissipative quiescent medium for planar propagation using the linearized compressible Navier-Stokes equation without the Stokes' hypothesis \cite{arxiv2023}, here the same is extended where a uniform mean flow is present in the ambiance to explore the propagation properties for the Doppler effect.   

\end{abstract}

\maketitle

\textbf{keywords:} fluid flow; acoustic field; dissipative medium; Doppler effect\\
\\
\section{Introduction}

For  problems of sound propagation, the use of the compressible Navier-Stokes equation is mandatory, with the disturbance treated as a small perturbation following the conservation of mass and momentum. In a flow with constant velocity, homogeneous medium, one can consider the equilibrium state with the following ansatz: disturbance velocity $\vec V '$ and disturbance density $\rho '$ develop with the constant mean flow ($\vec{ \bar V} = V_o \hat{i}$) and a steady state for the unperturbed density prevails i.e., $\left( \frac{\partial \bar \rho}{\partial t} = 0\right)$. The only assumption used here is the neglect of temperature perturbation following the small perturbation imposed on the pressure field, and this is the reason for not considering the conservation of energy equation. 

For the overall field, the conservation of mass is given by,
\begin{equation} \label{Eq:CE}
    \frac{\partial \rho}{\partial t} + \nabla \cdot \rho \vec{V} = 0 
\end{equation}
The corresponding conservation of momentum equation without any body force is given by,  
\begin{eqnarray}
&&\rho \left( \frac{\partial \vec V}{\partial t} + \left( \vec V \cdot \nabla\right) \vec V \right)  = -\nabla p  \nonumber \\
&&+ \nabla \cdot \left( \lambda \left(\nabla \cdot \vec V \right) \bf{I} \right) + \nabla \cdot \left[ \mu \left( \nabla \vec V + \nabla \vec V ^T \right)\right]  \label{Eq:ME}
\end{eqnarray}

Here, $\bf{I}$ is an identity matrix of rank three. If one considers the acoustic signal as a small perturbation over the mean flow, then the velocity, the density, and the pressure can be expressed as a superposition of the unperturbed equilibrium state with the small perturbation given by, $\vec V = \vec{\bar V} + \epsilon \vec V ' ; ~\rho     = \bar \rho + \epsilon \rho ' ;~ p = \bar p + \epsilon p '$. 
%

Without any loss of generality, one can consider the propagation of  disturbances over a constant mean flow (i.e., $\vec {\bar V} = V_o \hat{i}$) in a homogeneous medium (i.e. constant $\bar \rho$), so that, 
\begin{equation} \label{Eq:04}
\vec V = V_o \hat{i} + \epsilon \vec V'.
\end{equation}
The $O(\epsilon)$ equation resulting from the conservation of mass equation, Eq.~\eqref{Eq:CE} yields, 
\begin{equation} \label{Eq:05}
    \frac{\partial \rho '}{\partial t} + \bar \rho  \nabla \cdot \vec{V}' + V_o\frac{\partial \rho'}{\partial x} = 0 
\end{equation}	
Similarly, the $O(\epsilon)$ equation resulting from the conservation of momentum equation, Eq.~\eqref{Eq:ME} yields (with $\lambda$ and $\mu$ treated as constant in the absence of heat transfer), 

\begin{equation} \label{Eq:06}
\bar \rho  \frac{\partial \vec V '}{\partial t} + \bar \rho V_o\frac{\partial \vec V'}{\partial x} = 0   = -\nabla p' + \nabla \cdot  \left( \lambda \left(\nabla \cdot \vec V'  \right)\bf{I} \right) + \nabla \cdot \left[ \mu \left( \nabla \vec V' + \nabla \vec V^{' T} \right)\right]
\end{equation}
From Eq.~\eqref{Eq:05} one gets, $\nabla \cdot \vec V ' = -\frac{1}{\bar \rho} \frac{\partial \rho '}{\partial t} - \frac{V_o}{\bar \rho} \frac{\partial \rho'}{\partial x}$; and differentiating this equation with respect to time yields, 
\begin{equation} \label{Eq:07}
\frac{\partial}{\partial t} \left(  \nabla \cdot \vec V '\right) = -\frac{1}{\bar \rho} \frac{\partial^2 \rho'}{\partial t ^2} - \frac{V_o}{\bar \rho} \frac{\partial^2 \rho'}{\partial t \partial x}
\end{equation}
By taking divergence of Eq.~\eqref{Eq:06} one gets
\begin{equation} \label{Eq:08}
\bar \rho \frac{\partial}{\partial t} \left(  \nabla \cdot \vec V '\right) + \bar \rho V_o\left(  \nabla \cdot \vec V '\right) = -\nabla^2 p' + \lambda \nabla^2 \left( \nabla \cdot \vec V' \right) +  2\mu  \nabla^2 \left(\nabla \cdot \vec V'\right)
\end{equation}
Using Eq.~\eqref{Eq:07} in Eq.~\eqref{Eq:08} one gets, 
\begin{equation} \label{Eq:09}
-\frac{\partial^2 \rho'}{\partial t^2} - 2V_o\frac{\partial^2 \rho'}{\partial t \partial x} - V_o^2\frac{\partial^2 \rho'}{\partial^2 x}  = -\nabla^2 p' - \left( \lambda + 2 \mu\right) \left( \nabla^2 \frac{1}{\bar \rho} \frac{\partial \rho'}{\partial t} + \frac{V_O}{\bar \rho}\nabla^2\frac {\partial \rho'}{\partial x} \right) 
\end{equation}
Following the reasoning given in Ref. \onlinecite{arxiv2023}, One can invoke the polytropic relation between perturbation pressure and perturbation density (by neglecting temperature perturbation) as,

\begin{equation} \label{Eq:10}
	\frac{\partial \rho'}{\partial t} = \frac{1}{c^2}\frac{\partial p'}{\partial t} \quad \textrm{and} \quad  \frac{\partial \rho'}{\partial x} = \frac{1}{c^2}\frac{\partial p'}{\partial x} 
\end{equation}

In the absence of viscous losses, the medium's speed of sound ($c$) is for an isentropic process, such that one can treat a homogeneous temperature field for the propagation of other disturbances \cite{Feynman65,Feynman69}. Therefore, eliminating $\rho'$ using this relation in Eq.~\eqref{Eq:09}, one gets

\begin{equation} \label{Eq:11}
\frac{1}{c^2}\frac{\partial^2 p'}{\partial t^2} + \frac{2V_o}{c^2}\frac{\partial^2 p'}{\partial t \partial x} + \frac{V_o^2}{c^2}\frac{\partial^2 p'}{\partial^2 x}  =\nabla^2 p' + \frac{\lambda + 2 \mu}{\bar \rho c^2} \left(V_o\nabla^2 \frac{ \partial p'}{\partial x} + \nabla^2 \frac {\partial p'}{\partial t} \right) 
\end{equation}
\noindent which can be further simplified as,

\begin{equation} \label{Eq:12}
\frac{\partial^2 p'}{\partial t^2} + 2V_o \frac{\partial^2 p'}{\partial t \partial x} + V_o^2\frac{\partial^2 p'}{\partial^2 x} = c^2\nabla^2 p' + \nu_l \left(\frac{\partial }{\partial t} \nabla^2  p' + V_o \nabla^2 \frac{\partial p'}{\partial x}\right)
\end{equation}

\noindent where the generalized viscosity is defined as before \cite{AcousticPOF23} to be given by, $\nu_l = \frac{\lambda + 2 \mu}{\bar \rho}$, including the effects of the dissipative medium accounting for all forms of viscous losses.

\subsection{Characteristics of Perturbation Pressure Equation}

To elucidate the fundamentals of Doppler effects, attention is focused here in Eq.~\eqref{Eq:12} for the planar propagation of the disturbance field by considering a one-dimensional version of the perturbation equation, as it has been done for a quiescent ambiance \cite{Blackstock2000,AcousticPOF23} given by, 
\begin{equation} \label{Eq:13}
\frac{\partial^2 p'}{\partial t^2} + V_o^2\frac{\partial^2 p'}{\partial x^2} + 2V_o\frac{\partial^2 p'}{\partial x \partial t} - c^2\frac{\partial^2 p'}{\partial x^2} - \nu_l \left(V_o\frac{\partial^3  p' }{\partial x^3} + \frac{\partial^2 p'}{\partial t \partial x^2} \right) = 0
\end{equation}

For the purpose of dispersion and dissipation analysis, we represent the fluctuating pressure by, 
\begin{equation} \label{Eq:14}
p'(x,t) = \int \int \hat p (k,\omega)e^{ i (kx-\omega t)} dk~d\omega
\end{equation}
Rewriting Eq.\eqref{Eq:13} in the spectral plane by using the above representation in Eq.~\eqref{Eq:14}, one gets the quadratic dispersion relation as, 
\begin{equation} \label{Eq:15}
\omega ^2  -k \omega \left(2V_o - i \nu_l k \right) + V_ok^2 - i \nu_lk^3V_o - c^2 k^2 = 0 
\end{equation}
This yields the dispersion relation for the two components of the solution as 
\begin{equation}\label{Eq:16}
\omega_{1,2} = V_ok- \frac{i \nu_l k^2}{2} \pm kcf
\end{equation}
\noindent where we denote, $f = \sqrt{1-\left(\frac{\nu_l k}{2c} \right)^2}$. Treating the wavenumber $k$, as the independent variable, the dispersion relation in Eq.~\eqref{Eq:16}, provides the following amplification factors as \cite{Sengupta13,SumanSengPrasMoha17}, 
\begin{equation}
G_{1,2} = e^{-i \omega_{1,2} \tau_s}
\label{Eq:17}
\end{equation}
for the introduced time scale $\tau_s$. For $f > 0$, these complex exponents in Eq.~\eqref{Eq:17} the amplification factors are given for the real and imaginary parts for the first and second components by,

\begin{eqnarray}
G_{1, real} = e^{-k^2 \nu_l \tau_s/2} \cos k(V_o\tau_s + fc \tau_s) \\ 
G_{1, imag} = e^{-k^2 \nu_l \tau_s/2} \sin k(V_o\tau_s + fc \tau_s) \\ 
G_{2, real} = e^{-k^2 \nu_l \tau_s/2} \cos k(V_o\tau_s - fc \tau_s) \\
G_{2, imag} = e^{-k^2 \nu_l \tau_s/2} \sin k(V_o\tau_s - fc \tau_s) 
\label{Eq:18}
\end{eqnarray}
These real and imaginary parts of the amplification factors imposes phase shift for each components, and these can be obtained as given in the following \cite{AcousticPOF23},


\begin{equation} \label{Eq:19}
    \beta_{1,2} = V_o k \tau_s \pm k c f \tau_s, 
\end{equation}
By introducing the Mach number of the unperturbed flow as $M_o = V_o/c$, the non-dimensional phase speeds of the perturbation equation is given by, 
\begin{equation} \label{Eq:20}
    \frac{c_{ph 1,2}}{c} =\frac{\beta_{1,2}}{kc \tau_s} = M_o \pm f 
\end{equation}

The corresponding group velocity components ($v_{g 1,2}$), as given in the literature\cite{Rayleigh, Lighthill, Brillouin, Sengupta21}, of the perturbation equation are obtained from Eq.~\eqref{Eq:16} as, 

\begin{equation} \label{Eq:22}
    v_{g 1,2} = \frac{d \omega_{1,2}}{dk} = V_o \pm cf \mp \frac{\left(k \nu_l \right)^2}{4fc} - i \nu_l k
\end{equation}

This can be represented in the non-dimensional form as 

\begin{equation}\label{Eq:23}
\frac{v_{g 1,2}}{c} = M_0 \pm f \mp \frac{2}{f} (\frac{k}{k_c})^2 - \frac{ik\nu_l}{c}
\end{equation}

Where we have introduced the cut-off wavenumber as before \cite{AcousticPOF23,arxiv2023}, $k_c = 2c/\nu_l$. As has been explained before \cite{SumanSengPrasMoha17,AcousticPOF23}, the imaginary part of the group velocity is irrelevant for physical systems which do not exhibit anti-diffusion \cite{Sengupta13}, and will not be considered henceforth.

If the imaginary part of the physical amplification factor is absent in equation~\eqref{Eq:18}, then there will be no phase shift in the time interval of $\tau_s$. Such a situation can arise for $\beta_1 = m \pi$, for all integral values of $m$, including zero, i.e. 

\begin{eqnarray}
\label{Eq:24}
k_m\tau_s\left( V_o + cf_m \right)  = m\pi \;\;\; {\rm for}~m = 1,2, .... \infty \\
k_m\tau_s\left( V_o - cf_m \right)  = m\pi \;\;\; {\rm for}~m = 1,2, .... \infty 
\end{eqnarray}

\section{Results and Discussion}
\noindent We conduct an analysis utilizing Helium as the working medium with a temperature of $20^\circ$C. The corresponding values of density, speed of sound, and temperature are employed to calculate the cut-off wavenumber. To determine the generalized kinematic viscosity, the Stokes' hypothesis \citep{stokes1851} is applied, yielding $\nu_l = (4/3) \nu$. These properties are then plotted in the $(N_\tau, kL_s)$-plane based on this value of $\nu_l$. For the chosen working fluid, the speed of sound is calculated as $c = 1019.14 {\rm m/s}$ at an ambient temperature of $20^o$C and $\nu_l = 2.0 \times 10^{-4} {\rm m^2/s}$. This calculation yields a cut-off wavenumber of $k_c= 1.019 \times 10^{-7} {\rm m}^{-1}$. The maximum wavenumber ($k_{max}$) is set at four times the cut-off wavenumber. This choice allows us to determine the smallest resolved length-scale for the analysis, denoted as $L_s = 1.54012167 \times 10^{-7}$ m.

For all the figures presented in this section, the line representing the cut-off wavenumber $k_cL_s$ is displayed. As discussed in reference \citep{arxiv2023}, the solution's behavior distinguishes itself for wavenumbers above and below the defined cut-off wavenumber $k_c$ \citep{arxiv2023,AcousticPOF23}. Specifically, solutions with wavenumbers greater than $k_c$ exhibit a purely diffusive nature, while those with wavenumbers below $k_c$ display a wavy characteristic.

\begin{figure*}
\includegraphics[width=0.9\textwidth]{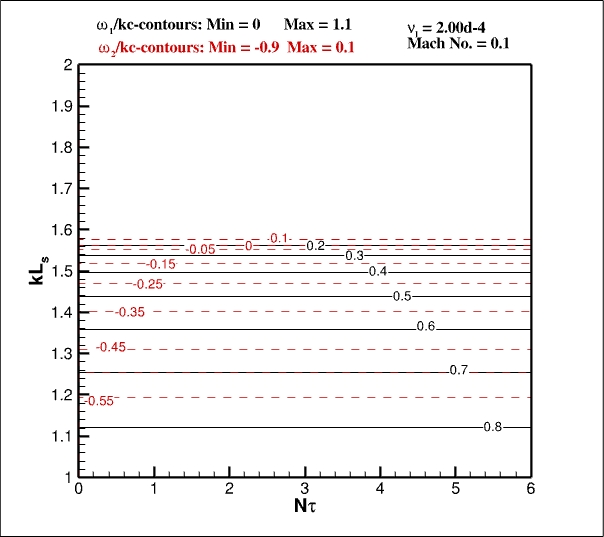}
\caption{The figure shows both modes of dispersion relation contours in the $(N_\tau, kL_s)$-plane, as given by equation (15) for Helium. The range of nondimensional wavenumber ($kL_s$) is fixed with $1.54012167 \times 10^{-7}$m fixed for the case of $\nu_l = 2.0 \times 10^{-4} {\rm m^2/s}$ and $M_o = 0.1$. The black line and red dotted line depict mode 1 and mode 2, respectively.}
\label{Fig1}
\end{figure*}

Figure \ref{Fig1} shows the contours of the real part of the dispersion relation mentioned in equation \eqref{Eq:16} for both modes. The Mach number for the imposed constant uniform flow is $0.1$. The generalized viscosity, $\nu_l$ is considered as $2 \times 10^{-4}$. Positive values represent the  waves whose phase gives it an appearance of a right-running wave and negative values seemingly represent the left-running wave. The first mode displays a higher maximum value of $\omega/kc$ because of the additional effect of constant uniform flow. This also resulted in the lower maximum value of the second mode compared to the first mode. 

\begin{figure*}
\includegraphics[width=0.9\textwidth]{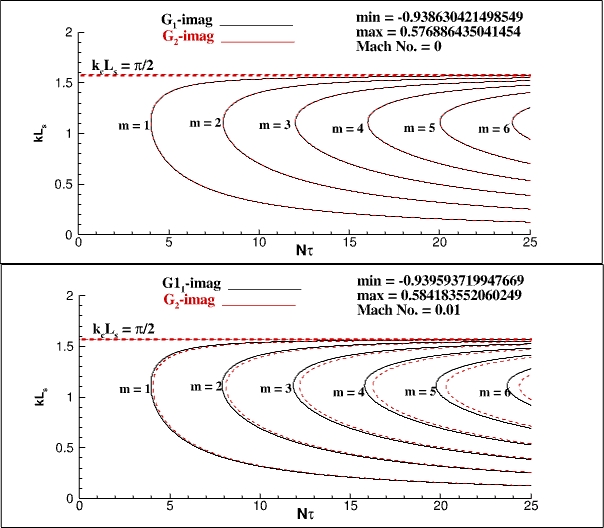}
\caption{The Frames show contours for both modes at which the imaginary part of the amplification factor will be zero in the $(N_\tau, kL_s)$-plane, as given by equation \eqref{Eq:18}. Top frame and bottom frame display the case of $\nu_l = 2.0 \times 10^{-4} {\rm m^2/s}$ for Mach number $M_o = 0$ and $M_o = 0.01$, respectively.}
\label{Fig2}
\end{figure*}

Figure \ref{Fig2} illustrates the zero contours of the imaginary part of the amplification factor, revealing a multi-modal behavior. The upper frame corresponds to a quiescent medium ($M_o = 0$), while the lower frame pertains to a medium with a constant uniform mean flow ($M_o = 0.01$). The first mode is represented by the black line, and the second mode is depicted by the red dotted line. The line $k_c L_s$ that separates the parabolic and hyperbolic regions is highlighted with a slightly thicker red dotted line across all the figures. If the imaginary part is absent in the amplification factor, there will be no phase shift. Observations reveal that multiple modes exhibit zero imaginary parts in the amplification factor. In the case of the quiescent medium, both modes of the imaginary part of the amplification factor overlap. However, a noticeable shift occurs in both modes under the influence of a mean flow. The first mode shifts to the left, while the second mode shifts to the right. This behavior can be explained by referencing equation \eqref{Eq:24}.

\begin{figure*}
\includegraphics[width=0.9\textwidth]{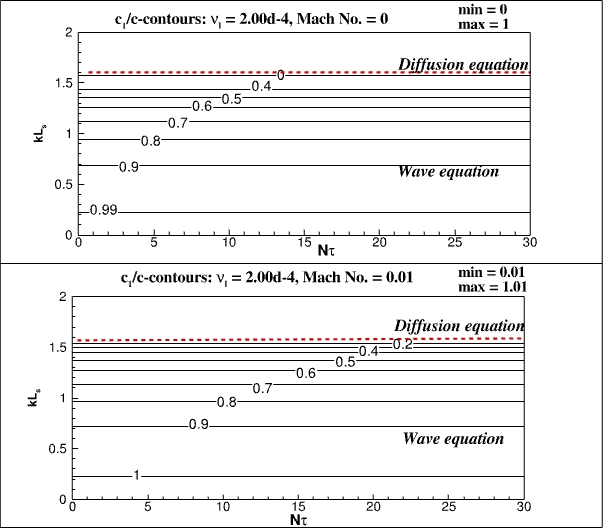}
\caption{The frames show the phase speed of first mode for the case of $\nu_l = 2.0 \times 10^{-4} {\rm m^2/s}$  Mach number $M_o = 0$ (Top frame) and $M_o = 0.1$ (Bottom frame) in the $(N_\tau, kL_s)$-plane, as given by equation~\eqref{Eq:20}. The length and time scales have been chosen as in figure~\ref{Fig1}. The boundary between the regions depicting the diffusion and wave solutions is given by the non-dimensional cut-off wavenumber ($k_c L_s$) shown by a dotted line (red).}
\label{Fig3}
\end{figure*}

Figure \ref{Fig3} illustrates the phase speeds of the first mode. The upper frame presents the phase speeds in a quiescent medium, while the lower frame displays the scenario with a constant mean flow. The phase speed of the first mode exhibits a non-trivial wavy solution for wave numbers ($k$) lower than a critical value ($k_c$). Conversely, when dealing with wave numbers above $k_c$, the governing equation assumes a diffusive nature, resulting in a constant phase over time due to the vanishing imaginary part of the amplification factor \cite{AcousticPOF23}. A notable observation in this figure is the discernible increase in the phase speed of the first mode under the influence of a constant mean flow, compared to the corresponding phase speed contour in the quiescent ambient flow scenario. The observed enhancement in the phase speed of the first mode can be attributed to the imposed flow conditions, as elucidated in Figure \ref{Fig1}.

\begin{figure*}
\includegraphics[width=0.9\textwidth]{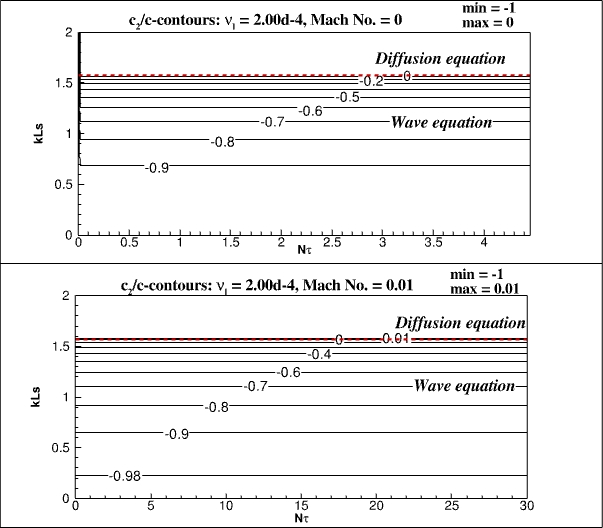}
\caption{The frames show the phase speed of second mode for the case of $\nu_l = 2.0 \times 10^{-4} {\rm m^2/s}$  Mach number $M_o = 0$ (Top frame) and $M_o = 0.01$ (Bottom frame) in the $(N_\tau, kL_s)$-plane, as given by equation~\eqref{Eq:20}. The length and time scales have been chosen as in figure~\ref{Fig1}. The boundary between the regions depicting the diffusion and wave solutions is given by the non-dimensional cut-off wavenumber ($k_c L_s$) shown by a dotted line (red).}
\label{Fig4}
\end{figure*}

Figure \ref{Fig4} presents the phase speed of the second mode. The upper frame pertains to the quiescent medium scenario, while the lower frame corresponds to the case with a constant mean flow. In the quiescent ambient situation, the phase speeds for both modes possess identical magnitudes but with opposite signs. In the presence of the imposed mean flow, the second mode's phase speed encompasses a range of values, extending from negative to positive. This observation suggests the simultaneous existence of both left-running and right-running phase. In contrast, Figure \ref{Fig3} demonstrated that the phase speeds of the first mode were strictly constrained to positive values. The actual propagation direction of wave components is determined by the group velocity, by which the energy propagates. 
\begin{figure*}
\includegraphics[width=0.9\textwidth]{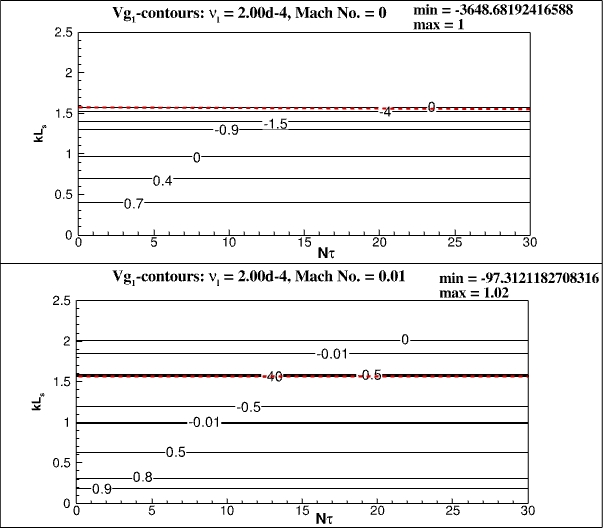}
\caption{The frames show group velocity of the first mode for the case of $\nu_l = 2.0 \times 10^{-4} {\rm m^2/s}$  Mach number $M_o = 0$ (Top frame) and $M_o = 0.01$ (Bottom frame) in the $(N_\tau, kL_s)$-plane, as given by equation~\eqref{Eq:23}. The length and time scales have been chosen as in figure~\ref{Fig1}. The boundary between the regions depicting the diffusion and wave solutions is given by the non-dimensional cut-off wavenumber ($k_c L_s$) shown by a dotted line (red).}
\label{Fig5}
\end{figure*}

Figure \ref{Fig5} illustrates the group velocity of the first mode. The upper frame pertains to the quiescent medium scenario, and the lower frame represents the case with a constant mean flow for $M_o = 0.01$. The group velocity plots reveal a multi-modal diffusive nature, as multiple zero crossing indicates propagation of disturbances in both directions with opposite signs of the group velocity present. Particularly noteworthy is the extension of the wave-like region beyond the cut-off wavenumber ($k_c$) for this case of a constant mean flow. There is the existence of multiple group velocity contours at the cut-off wavenumber for this non-quiescent condition.

\begin{figure*}
\includegraphics[width=0.9\textwidth]{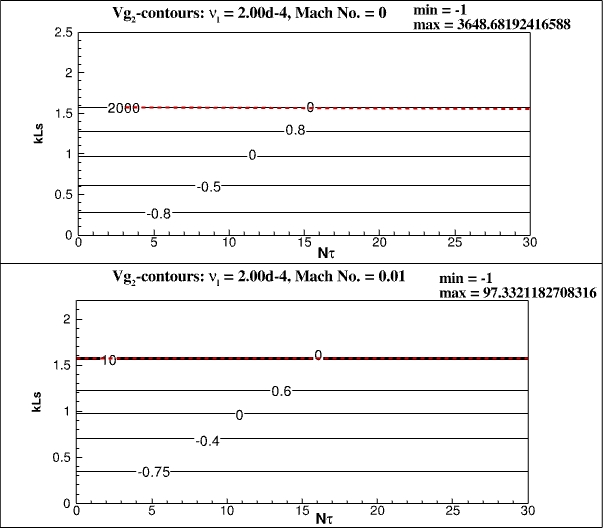}
\caption{The frames show group velocity of the second mode for the case of $\nu_l = 2.0 \times 10^{-4} {\rm m^2/s}$  Mach number $M_o = 0$ (Top frame) and $M_o = 0.01$ (Bottom frame) in the $(N_\tau, kL_s)$-plane, as given by equation~\eqref{Eq:23}. The length and time scales have been chosen as in figure~\ref{Fig1}. The boundary between the regions depicting the diffusion and wave solutions is given by the non-dimensional cut-off wavenumber ($k_c L_s$) shown by a dotted line (red).}
\label{Fig6}
\end{figure*}

Figure \ref{Fig6} depicts the group velocity of the second mode. The upper frame corresponds to the quiescent medium scenario, while the lower frame represents the case with a constant mean flow. The group velocity plots reveal a multi-modal diffusive nature. Unlike the first mode, there is no expansion of the wave-like region beyond the cut-off wavenumber $k_c$ in either scenario. Additionally, multiple group velocities are observed at the cut-off wavenumber for both quiescent and non-quiescent conditions. The occurrence of an extension to the wave-like region solely in the first case can be attributed to the positive constant velocity specified in the mean flow.

\section{Summary and Conclusions}
The perturbation field created by a pulse in quiescent ambiance has been reported recently \cite{AcousticPOF23}, where it has been shown that there exists a cut-off wavenumber ($k_c$), above which the governing wave equation for the perturbation pressure transforms to a diffusion equation, i.e. the acoustic wave diffuses the perturbation pressure. Furthermore, it has also been reported that such switching from wave equation to diffusion equation also occurs for sub-critical wavenumbers, explained as a multi-modal phenomenon \cite{arxiv2023}. All these significant results have been obtained from the governing equation of the perturbation field by including viscous diffusion, which was not included earlier, \cite{Feynman65,Feynman69,Whitham74}. The viscous loss terms were included for planar propagation in \cite{Blackstock2000,Trusler,Morse_Ingard}, but comprehensive details were communicated for the first time\cite{AcousticPOF23} for quiescent ambiance, by an analysis using global spectral analysis (GSA) \cite{Sagautetal2023,Sengupta13} to obtain the properties of the perturbation equation. GSA involves obtaining the metrics, such as the amplification factor, phase speed, and group velocity, by mapping the governing equation in the hybrid Fourier transform plane. The energy of a wave packet propagates with the group velocity, as noted \cite{Brillouin,Lighthill,Rayleigh} for electromagnetic waves, waves in fluids, and sound waves, respectively. Here, GSA has been extended in the presence of uniform mean flow, to quantify the well-known Doppler effect in acoustic. Detailed formulation of the perturbation equation has been provided here, with the uniform flow speed (or the Mach number, when non-dimensional quantities are presented), acting as a parameter. 

Unlike the isotropy of propagation for the wave equation for the case of quiescent ambiance, here the non-isotropy due to the Doppler effect is demonstrated by considering the case of planar propagation, as shown in Eq.~\eqref{Eq:16} for the dispersion relation. The corresponding plot is shown in Fig. \ref{Fig1}, and its effect on the amplification factor is shown in the nondimensional wavenumber and frequency plane, in Fig.~\ref{Fig2} for both the sub- and super-critical wavenumbers ranges. In Fig.~\ref{Fig2} and in all subsequent figures, the Doppler effect is demonstrated by comparing the quiescent case with a case for which the source is moving at a uniform speed given by the Mach number, $M_0 = 0.01$. All the figures are shown for a fixed generalized kinematic viscosity coefficient of $\nu_l= 0.0002$ $m^2/s$, typical of Helium as the working fluid with Stokes' hypothesis implemented. The Doppler effect is noted in the asymmetric attenuation for the left-running and right-running waves, as well as, for the modal morphing of the wave equation to diffusion equation for the sub-critical wavenumbers. 

The asymmetric theoretical properties are shown for Doppler effects by noting the phase speed and group velocities in Figs.~\ref{Fig3}-\ref{Fig4} and Figs.~\ref{Fig5}-\ref{Fig6}, respectively, where these are compared with the quiescent ambiance case. 

The study's implications extend to fluid dynamics and acoustics, offering insights into the propagation of sound, when it's source is in constant motion, both qualitatively and quantitatively for the first time. These findings contribute to a deeper understanding of the complex interplay between disturbances, flow conditions, and wave behavior in compressible fluid media. The results presented here contribute to advancing our knowledge precisely of wave phenomena \cite{Whitham74,Brillouin,Lighthill} and their practical implications for acoustics \cite{Blackstock2000,SenguptaBhum20}. It is proposed that the present research be extended to other types of mean flow fields, including for wall-bounded and free-shear layer flows.

\section*{AUTHOR DECLARATIONS}
\subsection*{Conflict of Interest}
The authors have no conflicts to disclose.

\section*{DATA AVAILABILITY}
The data that support the findings of this study are available from the corresponding author upon reasonable request.

\bibliography{acoustics_edited}
\end{document}